\RequirePackage{amsmath}

\documentclass[runningheads, dvipsnames]{llncs}
\usepackage[T1]{fontenc}
\usepackage{graphicx}

\usepackage{hyperref}
\usepackage{color}

\usepackage{amsmath}
\usepackage{amssymb}
\usepackage{mathtools}
\usepackage[noadjust]{cite}
\usepackage{enumitem}
\usepackage{caption}
\usepackage{subcaption}

\usepackage[linesnumbered, ruled, vlined]{algorithm2e}
\SetKwInOut{Input}{Input}
\SetKwInOut{Output}{Output}
\SetKw{Continue}{continue}
\SetKw{Break}{break}

\SetCommentSty{commentfont}
\DontPrintSemicolon

\newcommand\varcal[1]{\text{\usefont{OMS}{cmsy}{m}{n}#1}}
\newcommand{\bigO}{\varcal{O}}

\xspaceaddexceptions{]\}}

\newcommand{\nf}{\ensuremath{\mathit{\phi}}\xspace}
\newcommand{\singlenf}{\textsc{single-nf}\xspace}
\newcommand{\allnf}{\textsc{all-nf}\xspace}

\newcommand{\startpos}[1]{\ensuremath{\operatorname{start}(#1)}\xspace}
\newcommand{\str}[1]{\ensuremath{\operatorname{str}(#1)}\xspace}

\newcommand{\parent}[2][]{\ensuremath{\operatorname{parent}^{#1}(#2)}\xspace}
\newcommand{\depth}[1]{\ensuremath{\operatorname{depth}(#1)}\xspace}
\newcommand{\start}[1]{\ensuremath{\operatorname{start}(#1)}\xspace}
\newcommand{\child}[2]{\ensuremath{\operatorname{child}(#1, #2)}\xspace}
\newcommand{\children}[1]{\ensuremath{\operatorname{children}(#1)}\xspace}
\newcommand{\slink}[1]{\ensuremath{\operatorname{slink}(#1)}\xspace}
\newcommand{\wlink}[2]{\ensuremath{\operatorname{wlink}(#1, #2)}\xspace}
\newcommand{\wlinks}[1]{\ensuremath{\operatorname{wlinks}(#1)}\xspace}

\usepackage[capitalise, noabbrev]{cleveref}

\usepackage{xcolor}

\begin{document}
\title{Online Computation of String Net Frequency}
%
%
\author{
Peaker Guo\inst{1}\orcidID{0000-0002-9098-1783} \and
Seeun William Umboh\inst{1,3}\orcidID{0000-0001-6984-4007} \and
Anthony Wirth\inst{1,2}\orcidID{0000-0003-3746-6704} \and
Justin Zobel\inst{1}\orcidID{0000-0001-6622-032X}
}
\authorrunning{P.~Guo et al.}
%
\institute{
School of Computing and Information Systems, 
The University of Melbourne
\email{
zifengg@student.unimelb.edu.au,
william.umboh@unimelb.edu.au,
awirth@unimelb.edu.au,
jzobel@unimelb.edu.au
}
\and
School of Computer Science, The University of Sydney
\email{
anthony.wirth@sydney.edu.au
}
\and ARC Training Centre in Optimisation Technologies, Integrated Methodologies, and Applications (OPTIMA)
}
\maketitle              

\begin{abstract}
The \emph{net frequency} (NF) of a string, of length~$m$, in a text, of length~$n$, is the number of occurrences of the string in the text with unique left and right \emph{extensions}.
Recently, Guo et al.~[CPM 2024] showed that NF is combinatorially interesting and how two key questions can be computed efficiently in the \emph{offline} setting.
First, \singlenf: reporting the NF of a query string in an input text.
Second, \allnf: reporting an occurrence and the NF of each string of positive NF in an input text.
For many applications, however, facilitating these computations in an \emph{online} manner is highly desirable.
We are the first to solve the above two problems in the online setting, and we do so in optimal time, assuming, as is common, a constant-size alphabet: \singlenf in $\bigO(m)$ time and \allnf in $\bigO(n)$ time.
Our results are achieved by first designing new and simpler offline algorithms using suffix trees, proving additional properties of NF, and exploiting Ukkonen's online suffix tree construction algorithm and  results on \emph{implicit node} maintenance in an \emph{implicit suffix tree} by Breslauer and Italiano.

\keywords{
Suffix trees \and 
implicit nodes \and 
suffix links \and
Weiner links
}
\end{abstract}

\section{Introduction}

The \emph{net frequency} (NF) of a string~$S$, of length~$m$, in a text~$T$, of length~$n$, is 
the number of occurrences of~$S$ in~$T$ 
with unique left and right \emph{extensions}.
For example, let~$T$ be $\texttt{rstk\underline{st}castarstast\$}$.
Among the five occurrences of \texttt{st}, only the underlined occurrence 
has a unique left extension \texttt{kst} and a unique right extension \texttt{stc}.
In contrast, the other occurrences of \texttt{st} either have 
a repeated left extension (\texttt{rst} and \texttt{ast}),
or a repeated right extension (\texttt{sta}).
Thus the NF of \texttt{st} is 1 in $T$.
Introduced by Lin and Yu~\cite{journal/jise/2001/lin},
NF has been demonstrated to be useful
for Chinese phoneme-to-character (and character-to-phoneme) conversion 
and other NLP tasks~\cite{journal/jise/2001/lin, journal/ijclclp/2004/lin}.

Recently, Guo et al.~\cite{conf/cpm/2024/guo} reconceptualised NF and simplified the original definition.
They thus identified strings with positive NF, including
in Fibonacci words.
They also showed that there could be at most $n$ distinct strings in $T$ with positive NF.
They then bounded the sum of lengths of strings with positive NF in~$T$ of length~$n$ 
between $\Omega(n)$ and $\bigO(n \log \delta)$,
where $\delta := \max\left\{ S(k)/k : k \in [n] \right\}$ and~$S(k)$ is the number of distinct strings of length~$k$ in~$T$.

Although NF can be efficiently computed in the offline setting~\cite{conf/cpm/2024/guo}, 
it is also useful to determine NF in an online setting in
which a text is being dynamically extended.
In this setting, at all times we have read the first $n$ characters from a stream and further characters are to be read in turn, that is, $n$ is being incremented.
As each character arrives,
the data structure is updated and 
a query on NF computation can be answered.
Throughout our inspection of the online setting, $T$ is an input text of which we have read the first $n$ characters,
while $S$ is a length-$m$ query string. 
We consider the following two problems in the online setting.
$\underline{\singlenf}$: report the NF of $S$ in $T[1\ldots n]$.
$\underline{\allnf}$: for each string of positive NF in $T[1\ldots n]$, report one occurrence and its NF.

We have found that adapting the existing offline approach based on a suffix array
to the online setting is not trivial.
In the offline setting,
suppose we have read~$n$ characters of an input text~$T$ so far,
after building the structure for $T[1 \ldots n]$,
should we append a single character to~$T$, we would have to build the structure 
for $T[1 \ldots n+1]$ from scratch, without reusing the structure for  $T[1 \ldots n]$.
In contrast, in this paper,
the structure for $T[1 \ldots n+1]$ derives from the structure for $T[1 \ldots n]$.
This setting aligns with 
the \emph{online} suffix tree construction algorithm 
by Ukkonen~\cite{journal/algorithmica/1995/ukkonen},
which is widely used 
due to its simplicity and online nature, cited in numerous follow-up works~\cite{journal/jda/2013/breslauer, journal/tcs/2012/breslauer,conf/cpm/2006/inenage, conf/cpm/2014/larsson, conf/cpm/2016/takagi}.

\subsubsection{Our contribution.}
In this work, we first introduce a new characteristic of NF 
(\cref{thm:nf-char-suffix-tree}), 
specifically designed for NF computation with a suffix tree.
With this characteristic and Weiner links,
assuming a constant-size alphabet,
we present our optimal-time offline \singlenf algorithm 
(\cref{algo:single-nf-suffix-tree}).
Applying this characteristic in an asynchronous fashion,
our optimal-time offline \allnf algorithm 
(\cref{algo:all-nf-suffix-tree}) using suffix links
is arguably simpler than the state-of-the-art suffix array-based solution.
We then adapt our offline algorithms to the online setting -- to our knowledge online algorithms have not previously been reported.
With additional properties of NF and 
prior results on implicit nodes in an implicit suffix tree,
we obtain optimal $\bigO(m)$-time online \singlenf 
and $\bigO(n)$-time online \allnf algorithms
(\cref{thm:online-single-nf-result} and \cref{thm:online-all-nf-result}).

\section{Preliminaries}

\subsubsection{Strings.}
Let~$\Sigma$ be a constant-size alphabet throughout this paper. 
The constant-size assumption follows previous work on suffix trees~\cite{journal/tcs/2012/breslauer, conf/cpm/2014/larsson}.
Let $\mathit{SP} \equiv S \cdot P$ be the concatenation of two strings,~$S$ and~$P$.
Let~$[n]$ denote the set $\{1,2,\ldots,n\}$.
A substring of string~$T$ with starting position~$i \in [n]$ 
and end position~$j \in [n]$ is written as $T[i \ldots j]$. 
A substring $T[1\ldots j]$ is called a prefix of~$T$, 
while $T[i \ldots n]$ is called a suffix of $T$.
Let~$T_i$ denote the $i^{\text{th}}$ suffix of $T$, $T[i\ldots n]$.
An \emph{occurrence} in the text~$T$ is a pair of 
starting and ending positions $(i, j) \in [n] \times [n]$.
We say $(i, j)$ is an \emph{occurrence of string~$S$} if $S = T[i \ldots j]$,
and $i$ is an occurrence of $S$ if $S = T[i \ldots i + |S| - 1]$.
An occurrence $(i', j')$ is a \emph{sub-occurrence} of $(i, j)$
if $i \leq i'$, while $j' \leq j$.
The \emph{frequency} of~$S$ (in~$T$), denoted by~$f(S)$, 
is the number of occurrences of~$S$ in~$T$.

\subsubsection{Net frequency.}

The NF of a string was originally defined by Lin and Yu~\cite{journal/jise/2001/lin},
and reconceptualised and simplified by Guo et al.~\cite{conf/cpm/2024/guo}.
The NF of a unique string is defined to be zero;
the NF of a repeated string $S$, written as $\nf(S)$, 
is the number of occurrences of $S$ with unique left and right extensions.

\begin{definition}[Net occurrence~\cite{conf/cpm/2024/guo}]\label{def:net-occ}
An occurrence $(i, j)$ is a \emph{net occurrence} if
$f( T[i   \ldots j  ] ) \geq 2$, 
$f( T[i-1 \ldots j  ] ) = 1$, and
$f( T[i   \ldots j+1] ) = 1$.
When $i=1$, $f( T[i-1 \ldots j  ] ) = 1$ is assumed to be true;
when $j=n$, $f( T[i   \ldots j+1] ) = 1$ is assumed to be true.
\end{definition}

\noindent
NF can be also formulated in terms of symbols from $\Sigma$, 
rather than occurrences.

\begin{lemma}[NF characteristic~\cite{conf/cpm/2024/guo}]\label{thm:nf-char}
Given a repeated string $S$,
\[
\nf(S) =  \left| \left\{ \, 
(x,y)\in \Sigma \times \Sigma : f(xS)=1 \text{~and~}  f(Sy)=1 \text{~and~} f(xSy)=1 
\, \right\} \right| \,.
\]
\end{lemma}

\noindent
Here, strings $xS$, $Sy$, and $xSy$ are the  
\emph{left}, \emph{right}, and \emph{bidirectional extensions} of $S$, respectively.
Note that in~\cref{def:net-occ}, the condition 
on the bidirectional extension $f(T[i-1 \ldots j+1])=1$ is not needed because 
it is implied by $f( T[i-1 \ldots j  ] ) = 1$ and $f( T[i   \ldots j+1] ) = 1$.
A string $S$ is \emph{branching} in $T$ if  
$S$ is the longest common prefix of two distinct suffixes of $T$.
The following result says that branching strings are the only strings 
that could have positive NF.
\begin{lemma}[\cite{conf/cpm/2024/guo}]\label{thm:branching}
If $S$ is not a branching string, then $\nf(S)=0$.
\end{lemma}

\subsubsection{Suffix trees.}
Introduced by Weiner~\cite{conf/swat/1973/weiner},
the suffix tree is arguably one of the most significant and versatile 
data structures in string processing with a wide range of 
applications~\cite{book/cu/1997/gusfield}. 
The \emph{suffix tree} of $T$ is a rooted directed tree 
whose edges are labelled with substrings of $T$.
It contains $n$ leaf nodes, each labelled from $1$ to $n$.
For each $i \in [n]$, the concatenation of the edge labels 
on the path from the root to leaf $i$ forms the suffix $T_i$. 
Each non-root and non-leaf node, known as a \emph{branching node}, has at least two children,
and the first character of the label on the edge to each child is distinct.

It is commonly assumed that a unique sentinel character $\$ \notin \Sigma$ 
is appended to the text.
This ensures that no suffix is a prefix of another,
thereby ensuring that each suffix is represented as a distinct leaf in the suffix tree.
Given a node $u$, the \emph{path label} of $u$, denoted by $\str{u}$, 
is the concatenation of the edge labels 
on the path from the root to $u$.
We call $|\str{u}|$ the \emph{string depth} of $u$, denoted by $\depth{u}$.

In suffix-tree construction algorithms~\cite{journal/jacm/1976/mccreight, journal/algorithmica/1995/ukkonen, conf/swat/1973/weiner}, 
a type of pointer 
called a \emph{suffix link}\footnote{
A suffix link is also referred to as a \emph{vine pointer} in the
Prediction by Partial Matching algorithm~\cite{journal/tcom/1990/moffat} 
or as a \emph{failure link} in the
Aho--Corasick algorithm~\cite{journal/cacm/1975/aho}.
}
helps traverse the suffix tree efficiently; it is
a key ingredient in achieving linear time.
In our algorithms, besides suffix links,
we also use \emph{Weiner links}, 
also known as \emph{reverse suffix links}.
Given a branching node,~$u$, the \emph{suffix link} of~$u$
points from~$u$ to another branching node,~$v$,
where $\str{v}$ is the string obtained by removing the first character of~$\str{u}$, that is, the longest proper suffix of $\str{u}$.
Whenever there is a suffix link from $u$ to $v$,
there is a \emph{Weiner link}
from $v$ to $u$.
Suffix links and Weiner links are usually only needed among branching nodes,
but some applications also require them 
among leaf nodes~\cite{journal/jda/2015/starikovskaya}.

We now define the following notations in a suffix tree used in subsequent sections.
Consider a node $u$.
Let $\startpos{u}$ be the starting position of some occurrence of $\str{u}$ in the text.
Let $\parent{u}$ be the parent node of $u$.
Let $\child{u}{y}$ be the child node $v$ of $u$ such that 
the label of the edge $uv$ starts with symbol $y$,
and $\bot$ if such $v$ does not exist.
Let $\children{u} := \{ (v, y) : v = \child{u}{y} \neq \bot \}$.
Let $\slink{u}$ be the node that receives the suffix link from $u$.
Let $\wlink{u}{x}$ be the node $v$ that receives a Weiner link from $u$ and whose path label is $\str{v} = x \cdot \str{u}$, and $\bot$ if such $v$ does not exist.
Let $\wlinks{u} := \{ (v, x) : v = \wlink{u}{x}  \neq \bot \}$.

A suffix tree of $T$ can be also defined as a \emph{trie} for all the suffixes of $T$
where each non-branching path is compressed into a single edge. 
A \emph{locus} in a suffix tree corresponds to a node in the uncompressed trie.
More precisely,
it is a location in the suffix tree 
specified as a pair $(u, d)$ 
where $u$ is a node
and $d$ is an integer that satisfies $\depth{\parent{u}} < d \leq \depth{u}$.

Given a substring $S$ of $T$, after traversing the suffix tree 
from the root following the characters in~$S$,
the traversal always ends at a unique locus,~$(u, d)$. 
We say the traversal ends \emph{within} an edge if $d < \depth{u}$
and ends \emph{at} a node if $d = \depth{u}$.
Observe that the frequency of~$S$ equals the number of leaf nodes in the subtree rooted at~$u$.
So~$S$ is unique if~$u$ is a leaf node and 
is repeated if~$u$ is a branching node.
Also note that 
$S = \str{u}[1 \ldots d]$.

\subsubsection{Implicit suffix trees.}

The implicit suffix tree of a text can be defined by modifying its suffix tree as follows. 
First, every edge with only the sentinel character,~$ \$ $, as its label is removed.
Then, for each node~$u$ with only one child,~$v$, we perform these operations: let $w := \parent{u}$;
remove node $u$, together with  edges $wu$ and $uv$;
and create a new edge from $w$ to $v$ with edge label
$T[\start{v}+\depth{w} \ldots \start{v}+\depth{v}]$,
the concatenation of the edge labels of $wu$ and $uv$.
The tree obtained is called an \emph{implicit suffix tree}.
In an implicit suffix tree, branching nodes and leaf nodes are referred to as \emph{explicit nodes}.
Note that in a suffix tree, each suffix is unique and corresponds to a distinct leaf node.
However, in an implicit suffix tree, repeated suffixes
do not correspond to leaf nodes. 
For example, the repeated suffixes of the text \texttt{aabaabababaa}
are the strings \texttt{a}, \texttt{aa},  \texttt{baa}, and \texttt{abaa}.

\begin{definition}[Implicit node]
An \emph{implicit node} $(u, d)$ is a 
locus such that $\str{u}[1\ldots d]$ is a repeated suffix of the text.
\end{definition}

\begin{figure}[t]
\centering
\includegraphics[width=\linewidth]{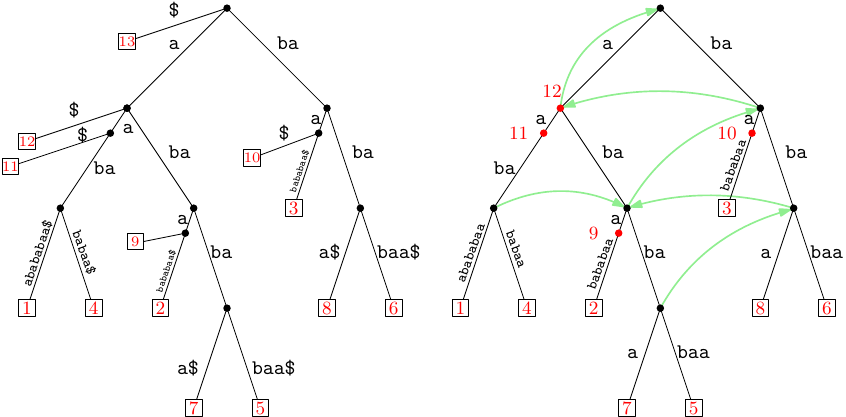}
\caption{
The suffix tree (left) and implicit suffix tree (right)
for text 
\texttt{aabaabababaa}.
Leaves (squares) and implicit nodes (red dots) are numbered;
green arrows are suffix links coming from branching nodes.
}
\label{fig:suffix-tree-examples}
\end{figure}

Each implicit node in an implicit suffix tree 
corresponds to a branching node in the suffix tree
with only two children and one of them is \$.
Note that if an implicit node satisfies $d = \depth{u}$,
then it coincides with a branching node in the implicit suffix tree.
\cref{fig:suffix-tree-examples} demonstrates the major aspects of the suffix tree and implicit suffix tree.
On its right, among the implicit nodes numbered~10--12,
node~12 coincides with a branching node.

The key idea of Ukkonen's algorithm is successive building 
of implicit suffix trees for each prefix of the text,
then adding the \$ at the end.
The algorithm maintains a pointer called the \emph{active point}
to the locus of the longest repeated suffix of the text.
One limitation of Ukkonen's algorithm is that
the locations of the implicit nodes are not maintained
during suffix tree construction.
Breslauer and Italiano~\cite{journal/tcs/2012/breslauer}
provide techniques for such task and extend Ukkonen's algorithm.
They also further classify edges and nodes in an implicit suffix tree.

\begin{definition}
An edge $uv$ is called an \emph{external edge} if $v$ is a leaf node
and an \emph{internal edge} otherwise.
An implicit node $u$ is called an \emph{internal implicit node}
if $u$ is within an internal edge and an \emph{external implicit node} otherwise.
\end{definition}

\noindent
If the locus each suffix corresponds to has an outgoing suffix link (including leaf and implicit nodes),
 then these suffix links 
form a path, which is called the \emph{suffix chain} by Breslauer and Italiano. 
The path starts from the leaf labelled by $T$ and 
ends at the root, going through each suffix and
satisfying the following.

\begin{lemma}[\cite{journal/tcs/2012/breslauer}]\label{thm:suffix-chain}
The suffix chain can be partitioned into the following consecutive segments: 
leaves, external implicit nodes, internal implicit nodes, and 
implicit nodes that coincide with branching nodes.
\end{lemma}

\noindent
On the right of \cref{fig:suffix-tree-examples},
these segments are: nodes 1--8, 9--10, 11, and 12.

The main result by Breslauer and Italiano that we use
together with Ukkonen's online suffix tree construction algorithm
is summarised as follows.

\begin{theorem}[\cite{journal/tcs/2012/breslauer, journal/algorithmica/1995/ukkonen}]\label{thm:implicit-node-query}
An implicit suffix tree on the first $n$ characters of $T$ , together with the suffix links and Weiner links can be built in $\bigO(n)$ amortized time.
A query returning the implicit nodes within a specific suffix tree edge takes worst-case $\bigO(1)$ time.
\end{theorem}

\noindent
Breslauer and Italiano showed that there is at most one implicit node within an internal edge.
Although each external edge may contain multiple implicit nodes, 
their representation as an arithmetic progression can be returned in $\bigO(1)$ time~\cite{journal/tcs/2012/breslauer}.
This result also supports other queries in the online setting~\cite{conf/cpm/2014/larsson}.

\section{Offline NF Computation with Suffix Trees}

We present our suffix-tree based approaches for offline NF computation,
which are arguably simpler than the suffix-array based approaches~\cite{conf/cpm/2024/guo}. 
We adapt them from the offline to the online setting in the following section.

\subsection{Offline \singlenf Algorithms}

The NF characteristic for suffix array might lead to 
explicit character matching in a suffix tree.
Moreover, when leaf nodes do not have incoming Weiner links, 
we are unable to enumerate unique left extensions of a string; 
so \cref{thm:nf-char} is unhelpful with a suffix tree.
We therefore introduce a new characteristic,
which is more suitable for computing NF in a suffix tree.
Essentially, it bypasses enumerating unique left extensions of the string.

\begin{theorem}[Suffix tree NF characteristic]\label{thm:nf-char-suffix-tree}
Given a repeated string $S$,
let $L(S) := \left \{ x \in \Sigma : f(xS) \geq 2 \right \}$
and $r(S) := \{ y \in \Sigma : f(Sy) = 1 \}$, then
\[
\nf(S) = |r(S)| - \sum_{x \in L(S)} | r(xS) \cap r(S) | \, . 
\]
\end{theorem}

\begin{proof}
We first define the following three sets:
\begin{align*}
L_{=1} &:= \left\{ (x,y)\in \Sigma \times \Sigma : f(xS) =1 \text{~and~} f(Sy) = 1 \text{~and~} f(xSy)=1 \right\},\\
L_{\geq 1} &:= \left\{ (x,y)\in \Sigma \times \Sigma : f(xS) \geq 1 \text{~and~} f(Sy) = 1 \text{~and~} f(xSy)=1 \right\}, \text{~and~}\\
L_{\geq 2} &:= \left\{ (x,y)\in \Sigma \times \Sigma : f(xS) \geq 2 \text{~and~} f(Sy) = 1 \text{~and~} f(xSy)=1 \right\}.
\end{align*}
Observe that $|L_{=1}| = |L_{\geq 1}| - |L_{\geq 2}|$ and
$\nf(S) = |L_{=1}|$.
Next, note that for each $y \in \Sigma$, 
if $f(Sy) = 1$, then
$f(xSy) \leq 1$ holds for each $x \in \Sigma$.
Further, if $f(xSy) = 1$ then $f(xS) \geq 1$;
otherwise, $f(xSy) = 0$ and $f(xS) = 0$.
Thus, we have
$|r(S)| = |L_{\geq 1}|$.
Finally, 
we derive that 
$
\sum_{x \in L(S)} |r(xS) \cap r(S)| 
= 
\left|  \bigcup_{x \in L(S)}  r(xS) \cap r(S) \right|
=
|L_{\geq 2}|.
$
Therefore, we have proved the desired result. 
\qed
\end{proof}
\noindent
Take~$T=\texttt{r\underline{st}k\underline{st}ca\underline{st}ar\underline{st}a\underline{st}\$}$~and~$S=\texttt{st}$ as an example.
We have $r(S) = \{ \texttt{\$}, \texttt{c}, \texttt{k} \}$ and 
$L(S) = \{ \texttt{r}, \texttt{a} \}$.
When~$x=\texttt{r}, r(xS) = \{ \texttt{a}, \texttt{k} \}$ 
and~$r(xS) \cap r(S) = \{ \texttt{k} \}$.
When~$x=\texttt{a}, r(xS) = \{ \texttt{\$}, \texttt{a} \}$
and~$r(xS) \cap r(S) = \{ \texttt{\$} \}$.
Thus, $\nf(S) = 3 - (1 + 1) = 1$.
The following informs checking of whether a right extension is unique.

\begin{proposition}\label{thm:unique-right-ext}
Consider a branching node $u$ and $y \in \Sigma$.
Let  $S := \str{u}$, then,~$Sy$ is unique if $\child{u}{y}$ is a leaf.
\end{proposition}

\cref{thm:nf-char-suffix-tree} immediately suggests a criterion
for early termination of the algorithm when it is determined that 
the NF of the query string is zero.

\begin{corollary}\label{thm:no-unique-child}
If $|r(S)|=0$, then $\nf(S)=0$.
\end{corollary}

\cref{thm:branching} was introduced to narrow down candidates
with a potentially positive NF when solving \allnf.
However, due to certain limitations of a suffix array, 
the result could not be efficiently applied to \singlenf. 
Now, with a suffix tree, 
we can utilise this result to detect a zero NF input string for \singlenf.

Our \singlenf algorithm using a  suffix tree 
is presented in \cref{algo:single-nf-suffix-tree}. 
It uses \cref{thm:branching} and \cref{thm:no-unique-child} for zero NF detection.
The algorithm computes the two terms in \cref{thm:nf-char-suffix-tree} separately and 
also utilises Weiner links for left extension enumeration.
The correctness of the algorithm follows from \cref{thm:nf-char-suffix-tree}.

Given a string $S$, \cref{algo:single-nf-suffix-tree} 
runs in $\bigO(|S|)$ time in the worst case.
Locating the locus of $S$ in the tree takes $\bigO(|S|)$ time.
Computing $|r(S)|$ takes at most $\bigO(|\Sigma|)$ time,
a constant in our analysis.
Then the time to compute $\sum_{x \in L(S)} | r(xS) \cap r(S) |$
is bounded by 
$|\Sigma|^2$,
also a constant.

\begin{algorithm}[t]
\caption{for offline \singlenf}
\label{algo:single-nf-suffix-tree}
$(u, d) \gets $ the locus of a query string $S$;\;
\tcp*[h]{unique or non-branching strings have zero NF (\cref{thm:branching})}\;
\lIf{$u$ is a leaf or $d < \depth{u}$}{\Return 0;}

\tcp*[h]{initialise the NF of $S$ by counting $|r(S)|$ in \cref{thm:nf-char-suffix-tree}}\;
$\nf \gets $ the number of leaf child nodes of $u$; \;

\tcp*[h]{if $|r(S)|=0$, then $\nf(S)=0$ (\cref{thm:no-unique-child})}\;
\lIf{$\nf=0$}{\Return 0;}

\tcp*[h]{compute $\sum_{x \in L(S)} | r(xS) \cap r(S) |$ in \cref{thm:nf-char-suffix-tree}
}\;
\ForEach{$(v, x) \in \wlinks{u}$}{ 
    \ForEach{$(w, y) \in \children{v}$ }{  
        \tcp*[h]{$xSy$ and $Sy$ are both unique (\cref{thm:unique-right-ext})}\;
        \lIf{$w$ is a leaf and $\child{u}{y}$ is a leaf}{
            $\nf \gets \nf - 1$;
        }
    }
}
\Return $\nf$;
\end{algorithm}

\subsection{Offline \allnf Algorithm}

From \cref{thm:branching}, only branching strings 
could have positive NF.
In this section, we present an offline \allnf algorithm that 
extracts and stores the NF of each branching string
in its corresponding branching node in the suffix tree.
The stored positive NF values can be reported afterwards.

In \cref{algo:single-nf-suffix-tree}, 
the two terms in in \cref{thm:nf-char-suffix-tree} 
are computed in separate steps of the algorithm.
For our \allnf algorithm, we compute these two terms 
in an \emph{asynchronous} fashion:
before we finish computing the NF of one branching string,
we might start computing the NF of another branching string.
In other words, the NF of each branching string 
is partially computed and updated as the algorithm progresses.
Specifically, suppose we are visiting a branching node $v$, 
let $u := \slink{v}$,
$xS := \str{v}$, and $S := \str{u}$.
We update the NF for both $xS$ and $S$ as follows.
We compute $|r(xS)|$ to update $\nf(xS)$ and 
compute $| r(xS) \cap r(S) |$ to update $\nf(S)$.
Note that, after visiting node $v$,
it is possible that neither $\nf(xS)$ nor $\nf(S)$
has the correct value, as they may have only been partially computed. 
But at the end of the algorithm, each NF value will be correct.

\begin{figure}[t]
\begin{minipage}{0.62\textwidth}

\begin{algorithm}[H]
\caption{for offline \allnf}
\label{algo:all-nf-suffix-tree}
\tcp*[h]{
assume $\nf(\str{v})$ is initialised to 0 for each branching node $v$ in the suffix tree
}\;
\ForEach{branching node $v$ in the suffix tree}{
    $u \gets \slink{v}; \ xS \gets \str{v} ; \ S \gets \str{u}; $\;
    \ForEach{$(w, y) \in \children{v}$}{
        \If{$w$ is a leaf}{
            \tcp*[h]{$xSy$ is unique}\;
            $\nf(xS) \gets \nf(xS) + 1$;\;
            \If{$\child{u}{y}$ is a leaf}{
                \tcp*[h]{$Sy$ is unique}\;
                $\nf(S) \gets \nf(S) - 1$;
            }
        }
    }
}
\end{algorithm}

\end{minipage}
\hfill
\begin{minipage}{0.38\textwidth}

\centering
\includegraphics[width=0.9\linewidth]{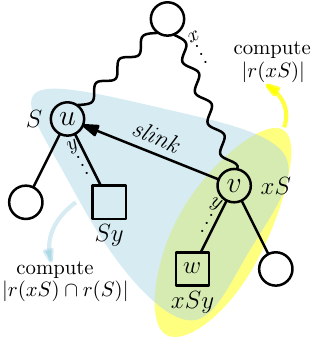} 
\caption{
Algo. \ref{algo:all-nf-suffix-tree} illustration.
}
\label{fig:all-nf-illustration}

\end{minipage}
\end{figure}

Our offline \allnf algorithm 
is listed in \cref{algo:all-nf-suffix-tree}
and illustrated in  \cref{fig:all-nf-illustration}.
It does not require traversing
the branching nodes in a particular order, and
runs in $\bigO(n)$ time 
in the worst case.
Since checking whether a node is a leaf takes $\bigO(1)$ time,
the overall time usage is bounded by the number of nodes 
visited throughout execution. 
Given a branching node,~$u$,
each of its child nodes,~$w$, is visited exactly once; 
each node in the suffix tree is 
the child node of exactly one branching node.
Moreover, there are at most $n$ branching nodes in the tree.
Thus, the total number of nodes visited is bounded by $\bigO(n)$.

\section{Online NF Computation with Implicit Suffix Trees}

With our offline algorithms, we first introduce additional properties
of NF then present our online algorithms based on these results.

\subsection{Online \singlenf Algorithm}

Our offline \singlenf algorithm 
computes the two terms in \cref{thm:nf-char-suffix-tree} 
for a query string whose locus is a branching node.
In the online setting, our online approach is still based on \cref{thm:nf-char-suffix-tree},
but also taking implicit nodes into account in the implicit suffix tree.
We first reconsider \cref{thm:unique-right-ext}, as \cref{thm:unique-right-ext-implicit}.

\begin{proposition}\label{thm:unique-right-ext-implicit}
Consider a branching node,~$u$, 
and $y \in \Sigma$.
Let $S := \str{u}$, then,~
$Sy$ is unique if $v := \child{u}{y}$ is a leaf and edge~$uv$ has no implicit node.
\end{proposition}

Let $(u, d)$ be the locus of the query string $S$.
In the offline case, $|r(S)|$ in \cref{thm:nf-char-suffix-tree}  
is computed by simply counting the number of leaf child nodes of $u$.
But it is more involved in the online setting.

\begin{figure}[t]
\centering
\includegraphics[width=0.8\linewidth]{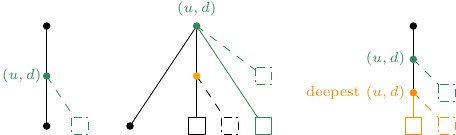}
\caption{Illustration of \cref{thm:locus-unique-right-ext}:
\labelcref{case-1} (left), \labelcref{case-2} (middle), and \labelcref{case-3} (right).
Black dots, coloured dots, and squares represent branching nodes, implicit nodes, and leaves, respectively.
Each dashed (non-existent) edge has label~\$
and leads to a dashed (non-existent) leaf node.
In each case, 
each implicit node has its own colour.
The implicit node $(u, d)$ is also labelled,
and the leaf nodes corresponding to its unique right extensions 
share the same colour as $(u, d)$.
}
\label{fig:implicit-right-ext}
\end{figure}

\begin{definition}
Consider an implicit node $(u, d)$ and
let $\rho(u, d)$ denote the number of unique right extensions of $\str{u}[1 \ldots d]$.
\end{definition}

\begin{lemma}\label{thm:locus-unique-right-ext}
Consider an implicit node $(u, d)$ and let $S := \str{u}[1 \ldots d]$.
The following cases are illustrated in \cref{fig:implicit-right-ext}.
\begin{enumerate}[label=Case \arabic*]
\item\label{case-1} If $(u, d)$ is an internal implicit node, then $\rho(u, d) = 1$.
\item\label{case-2} If $(u, d)$ is an internal implicit node that coincides with a branching node,
then $\rho(u, d)$ equals one plus the number of leaf child nodes of~$u$
whose leading edge does not contain an implicit node.
\item\label{case-3} If $(u, d)$ is an external implicit node,
then  $\rho(u, d) = 2$ if $(u, d)$ is the deepest such on the edge,
and $\rho(u, d) = 1$ otherwise.
\end{enumerate} 
\end{lemma}

\begin{proof}
In \labelcref{case-1}, the only unique right extension character is the \$.
\labelcref{case-2} follows from \cref{thm:unique-right-ext-implicit}, and we add one for the \$.
In \labelcref{case-3}, the deepest such has the \$ and a leaf child, while the others only has the \$.
\qed
\end{proof}

\noindent
The NF of the longest repeated suffix is given as follows.
\begin{lemma}\label{thm:nf-longest-repeated-suffix}
Consider an implicit node $(u, d)$ and let $S := \str{u}[1 \ldots d]$.
If $S$ is the longest repeated suffix of $T$,
then $\nf(S) = \rho(u, d)$.
\end{lemma}
\begin{proof}
Observe that each occurrence of $S$ has a unique left extension character as otherwise $S$ would not be the longest repeated suffix.
So, $\nf(S) = \rho(u, d)$.
\qed
\end{proof}
The following result provides another zero NF detection mechanism. 

\begin{lemma}\label{thm:implicit-node-zero-nf}
Consider an implicit node $(u, d)$ and let $S := \str{u}[1 \ldots d]$.
If $S$ is not the longest repeated suffix and 
$(u, d)$ does not coincide with a branching node,
then $\nf(S) = 0$.
\end{lemma}

\begin{proof}
First consider each occurrence of $S$ that is 
not a sub-occurrence of an occurrence of the longest repeated suffix:
the left extension character of such occurrence is always $T[n-d]$.
Next, consider each occurrence of $S$ except the rightmost one:
since $(u, d)$ does not coincide with a branching node,
the right extension character of such occurrence is always $T[i+d]$
for any occurrence $i$ of $S$ with $i < n-d$.
Thus, no occurrence of $S$ is a net occurrence and $\nf(S) = 0$.
\qed
\end{proof}

\begin{figure}[t]
\centering
\begin{subfigure}{0.49\columnwidth}
    \centering
    \includegraphics[width=\linewidth]{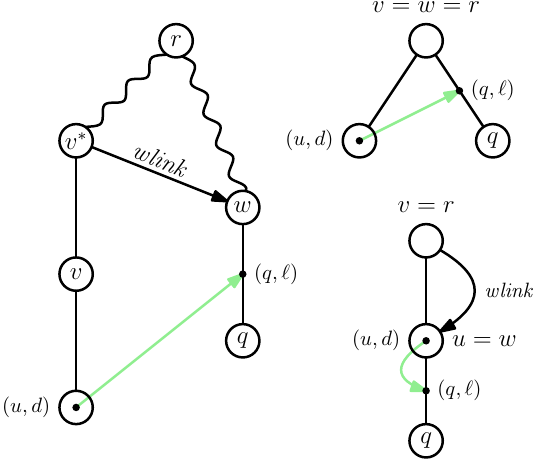}
    \caption{Illustration of  \cref{thm:wlink}.}
\label{fig:wlink-theorem}
\end{subfigure}
\hfil
\begin{subfigure}{0.45\columnwidth}
    \centering
    \includegraphics[width=\linewidth]{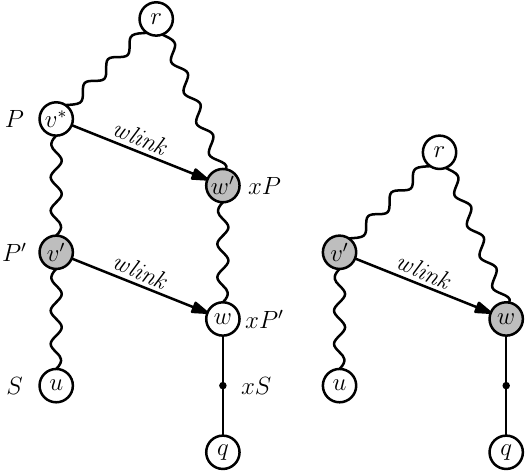}
    \caption{Illustration of proof of \cref{thm:wlink}.}
    \label{fig:wlink-proof}
\end{subfigure}
\caption{
Let~$r$ be the root node of the implicit suffix tree. 
An edge is shown straight; a path is shown squiggly.
In \cref{fig:wlink-theorem}, some possible locations of 
nodes~$u$, $w$, and $v := \parent{u}$ are shown:
$v \neq r \neq w$ (left),
$v = r = w$ (top right),
$v= r \neq u = w$ (bottom right).
Each green arrow indicates an implicit Weiner link
from~$(u, d)$ to~$(q, \ell)$.
In \cref{fig:wlink-proof}, compare scenarios
when $v^*$ exists (left) and when $v^*$ does not exist (right).
A node is coloured grey to indicate that it exists only under false assumption.
Next to several nodes are corresponding path labels.
}
\end{figure}

For the remaining case, not covered by \cref{thm:nf-longest-repeated-suffix} 
or \cref{thm:implicit-node-zero-nf},
the NF of the query string cannot be easily deduced.
Further computation, assisted by Weiner links, is required.
In \cref{algo:single-nf-suffix-tree},
the recipient of each Weiner link is a branching node,
however, for online \singlenf, \emph{implicit} Weiner links, 
whose recipients are implicit nodes, are also needed. 
We present the following result on how to compute an implicit Weiner link.
The result is illustrated in \cref{fig:wlink-theorem}.

\begin{lemma}\label{thm:wlink}
Consider an implicit node $(u, d)$ 
that coincides with a branching node.
Let $(q, \ell)$ be another implicit node such that 
$\str{q}[1 \ldots \ell] = x \cdot \str{u}[1 \ldots d]$
for some $x \in \Sigma$.
Define $w := \parent{q}$.
If there exists an ancestor $v$ of $u$ such that 
$\wlink{v}{x}$ exists, then
$w = \wlink{v^*}{x}$
where $v^*$ is the lowest such ancestor of $u$;
otherwise, $w$ is the root.
\end{lemma}

\begin{proof}
The proof is illustrated in \cref{fig:wlink-proof}.

If $v^*$ exists,
let $w' := \wlink{v^*}{x}$ and 
assume, by contradiction, that $w' \neq w$.
Let $P := \str{v^*}$, then $\str{w'} = xP$.
Since $xP$ is a prefix of $xS$,
$w'$ is an ancestor of $w$.
Let $P' := \str{w}[2 \ldots \depth{w}]$,
then $xP' = \str{w}$.
Since $w$ is a branching node,
there exists a branching node $v'$ with $\str{v'} = P'$.
This implies that $\wlink{v'}{x} = w$.
Since $v'$ is an ancestor of $u$,
this contradicts that $v^*$ is the lowest such ancestor.
Thus, our assumption is false and $w = w'$.

If $v^*$ does not exist,
assume, by contradiction, that $w$ is not the root.
Similarly, since $w$ is a branching node,
there exists a branching node $v'$ such that $\wlink{v'}{x} = w$.
Since $v'$ is an ancestor of $u$,
this contradicts that $v^*$ does not exist.
Thus, our assumption is false and $w$ is the root.
\qed
\end{proof}

\noindent
Note that the concept of implicit Weiner links has also been described before~\cite{conf/cpm/2019/belazzougui, journal/tcs/2023/fujishige, journal/tcs/2022/nakashima},
but with either a different definition or way of computing them.
We next introduce a corollary of \cref{thm:nf-char-suffix-tree}
that splits the set $L(S)$ into two disjoint sets.

\begin{corollary}\label{thm:nf-char-split-l}
Consider a locus $(u, d)$ with $d = \depth{u}$. 
Let $S := \str{u}$, 
\begin{align*}
L_{\mathit{exp}}(S) &:=
\{ x \in \Sigma : f(xS) \geq 2 \text{~and the locus of~} xS \text{~is an explicit node} \}, \\
L_{\mathit{imp}}(S) &:=
\{ x \in \Sigma : f(xS) \geq 2 \text{~and the locus of~} xS \text{~is an implicit node} \},
\end{align*}
and let
$\lambda(x) := | r(xS) \cap r(S) | $, then
\[
\nf(S) = |\rho(u, d)| 
- \sum_{x \in L_{\mathit{exp}}(S)} \lambda(x) 
- \sum_{x \in L_{\mathit{imp}}(S)} \lambda(x) \, . 
\]
\end{corollary}

Using \cref{thm:wlink} and \cref{thm:nf-char-split-l},
together with the results introduced in this section earlier,
we present our online \singlenf algorithm in 
\cref{algo:single-nf-implicit}.
Computing implicit Weiner links takes $\bigO(|S| \cdot |\Sigma|^2) \subseteq \bigO(|S|)$ time, assuming a constant-size alphabet.
The rest of the analysis follows from the analysis of \cref{algo:single-nf-suffix-tree}
and we have the following result.

\begin{theorem}\label{thm:online-single-nf-result}
Online \singlenf can be solved in worst-case $\bigO(|S|)$ time.
\end{theorem}

\begin{algorithm}[t]
\caption{for online \singlenf
}
\label{algo:single-nf-implicit}
$(u, d) \gets $ the locus of a query string $S$;\;

\lIf{$(u, d)$ is the active point}{\Return $\rho(u, d)$; 
\tcp*[h]{\cref{thm:nf-longest-repeated-suffix}} 
}

\lIf{$u$ is a leaf or $d < \depth{u}$}{\Return 0;
\tcp*[h]{\cref{thm:branching} and \cref{thm:implicit-node-zero-nf}}
}

$\nf \gets \rho(u, d)$; 
\tcp*[h]{initialise the NF of $S$ by computing $\rho(u, d)$ 
using \cref{thm:locus-unique-right-ext}}
\;

\lIf{$\nf=0$}{\Return 0;
\tcp*[h]{\cref{thm:no-unique-child}}
}

\tcp*[h]{compute $\sum_{x \in L_{\mathit{exp}}(S)} \lambda(x)$ 
in \cref{thm:nf-char-split-l}}\;
\ForEach{$(w, x) \in \wlinks{u}$}{
    \ForEach{$(q, y) \in \children{v}$ }{
        \lIf{$q$ is a leaf and 
             edge $vq$ does not contain an implicit node and 
             $\child{u}{y}$ is a leaf
             }{
            $\nf \gets \nf - 1$;
        }
    }
}

\lIf{$(u, d)$ is not an implicit node}{\Return $\nf$;}

\tcp*[h]{
compute $\Lambda$, the set of implicit Weiner links from $(u, d)$
using \cref{thm:wlink} 
}
\;
$\Lambda \gets \emptyset; \ r \gets \text{the root}; \ p \gets u$;\;

\While{$p \neq r$}{
    $p \gets  \parent{p}$;\;
    \ForEach(\tcp*[h]{``\_'' is used when the character is unused}){$(w, \_) \in \wlinks{p}$}{
        \ForEach{$(q, \_) \in \children{w}$ }{
            \lIf{there exists implicit node $(q, \ell)$ with $\ell = d+1$}{
                $\Lambda \gets \Lambda \cup \{ (q, \ell) \}$;
            }
        }
    }
}

\ForEach{$(q, \_) \in \children{r}$ }{
    \lIf{there exists implicit node $(q, \ell)$ with $\ell = d+1$}{
        $\Lambda \gets \Lambda \cup \{ (q, \ell) \}$
    }
}

\tcp*[h]{compute $\sum_{x \in L_{\mathit{imp}}(S)} \lambda(x)$ 
in \cref{thm:nf-char-split-l};
note that each $\str{q}[1\ldots \ell]$ has exactly two right extension characters, \$ (assumed to be unique) and $T[\startpos{q} + \ell +1]$}\;

\ForEach{$(q, \ell) \in \Lambda$}{
    $\nf \gets \nf - 1$; \tcp*[h]{for \$ 
} \;
    $y \gets T[\startpos{q} + \ell +1]$;\;
    \lIf{$q$ is a leaf and $\child{u}{y}$ is a leaf}{
        $\nf \gets \nf - 1$;
    }
}

\Return $\nf$;
\end{algorithm}

\subsection{Online \allnf Algorithm}

Our arguably very simple offline \allnf algorithm 
forms the basis of our online approach.
Similar to the online \singlenf algorithm,
we need to deal with left extensions whose loci are implicit nodes.
We make an additional observation.

\begin{lemma}
\label{lem:chainreverse}
Consider a repeated suffix~$S$ and its left extension,~$xS$,
which is also a repeated suffix.
If the locus of~$xS$ coincides with a branching node,
then the locus of~$S$ also does.
If the locus of~$S$ does not coincide with a branching node,
then the locus of~$xS$ also does not.
\end{lemma}
\begin{proof}
Consider the suffix chain in \cref{thm:suffix-chain} in reverse
and we have the desired result.
\qed
\end{proof}
With \cref{lem:chainreverse}, observe that at most one string might require 
implicit Weiner links: namely, the longest repeated suffix whose locus
coincides with a branching node.
The other repeated suffixes whose locus coincides with a branching node
do not have implicit Weiner links.
Our online \allnf algorithm is adapted from our offline \allnf algorithm as follows.
We first compute the NF of the longest repeated suffix.
We then invoke offline \allnf, but use \cref{thm:unique-right-ext-implicit} 
instead of \cref{thm:unique-right-ext}. 
While traversing each branching node,
we keep track of string depth and find~$\tau$,
the longest repeated suffix whose locus coincides with a branching node.
Once we do, we invoke online \singlenf on $\tau$.
Overall, the cost is bounded by $\bigO(n + |\tau|) \subseteq \bigO(n)$ where 
$\bigO(n)$ is the cost for offline \allnf and
$\bigO(|\tau|)$ is the cost for online \singlenf on $\tau$.
\begin{theorem}\label{thm:online-all-nf-result}
Online \allnf can be solved in worst-case $\bigO(n)$ time.
\end{theorem}

\section{Conclusion and Future Work}

In this work, we present, to our knowledge,
the first and, indeed, optimal-runtime online algorithms 
for both \singlenf and \allnf.
Having unsuccessfully investigated online approaches based
on suffix arrays -- the basis of previous offline
methods -- we turned our attention to suffix trees,
using which we found offline and online
algorithms that are runtime-optimal.
The results are based on new characteristics and properties of net frequency
and prior results on auxiliary pointers in suffix trees
and on implicit node maintenance. 

An avenue of future work is design and engineering an efficient implementation
of our solution, in particular the structures required for 
implicit nodes such as structures for dynamic nearest marked ancestors in trees~\cite{conf/cpm/2014/feigenblat, journal/iandc/1985/amir}
and structures for ancestor relationships~\cite{conf/esa/2002/bender}.

\begin{credits}
\subsubsection{\ackname} 
The authors thank Patrick Eades for insightful discussions during the early stage of this work.
The authors also
thank the anonymous reviewers for their suggestions.
This work was supported by the Australian Research Council, 
grant number DP190102078, 
and an Australian Government Research Training Program Scholarship.
\end{credits}

%
%
%
\bibliographystyle{splncs04}
\bibliography{references}

\end{document}